\documentclass[a4paper,twocolumn,11pt,accepted=2020-09-15]{quantumarticle}
\pdfoutput=1
\usepackage[utf8]{inputenc}
\usepackage[english]{babel}
\usepackage[T1]{fontenc}
\usepackage[usenames,dvipsnames]{xcolor}
\usepackage{amsmath}
\usepackage[colorlinks=true,citecolor=Blue,linkcolor=Blue,urlcolor=Blue]{hyperref}
\usepackage[numbers,sort,compress]{natbib}
\usepackage{tikz}
\usepackage{lipsum}

\usepackage{times,graphics,graphicx,bm,bbm,bbold,amssymb,amsfonts,dsfont,cancel,color}
\usepackage{latexsym}
\DeclareFontFamily{OT1}{pzc}{}
\DeclareFontShape{OT1}{pzc}{m}{it}
              {<-> s * [1.25] pzcmi7t}{}
\DeclareMathAlphabet{\mathpzc}{OT1}{pzc}
                                 {m}{it}
\usepackage[T1]{fontenc} 

\newcommand\blfootnote[1]{%
\begingroup
\renewcommand\thefootnote{}\footnote{#1}%
\addtocounter{footnote}{-1}%
\endgroup
}

\def\la{\langle}
\def\ra{\rangle}
\def\om{\omega}
\def\Om{\Omega}

\def\tr{{\rm Tr}}

\newcommand{\beq}{\begin{equation}}
\newcommand{\eeq}{\end{equation}}
\newcommand{\beqa}{\begin{eqnarray}}
\newcommand{\eeqa}{\end{eqnarray}}

\newcommand{\ket}[1] {\vert #1 \rangle}
\newcommand{\bra}[1] {\langle #1 |}

\newcommand{\Tr}{{\rm Tr}}

\begin{document}
\sloppy
\title{Shortcuts to Adiabaticity in Driven Open Quantum Systems: 
Balanced Gain and Loss and Non-Markovian Evolution}

\author{S. Alipour$^*$}
\affiliation{QTF Center of Excellence, Department of Applied Physics, Aalto University, P. O. Box 11000, FI-00076 Aalto, Espoo, Finland}

%\orcid{}

\author{A. Chenu$^*$}
\affiliation{Donostia International Physics Center,  E-20018 San Sebasti\'an, Spain}
\affiliation{IKERBASQUE, Basque Foundation for Science, E-48013 Bilbao, Spain}
\orcid{0000-0002-4461-8289}

\author{A. T. Rezakhani}
\affiliation{Department of Physics, Sharif University of Technology, Tehran 14588, Iran}
%\orcid{}

\author{A. del Campo}
\affiliation{Donostia International Physics Center,  E-20018 San Sebasti\'an, Spain}
\affiliation{IKERBASQUE, Basque Foundation for Science, E-48013 Bilbao, Spain}
\affiliation{Department of Physics, University of Massachusetts, Boston, MA 02125, USA}
\orcid{0000-0003-2219-2851}

\maketitle

%%%%%%%%%%%%%%%%%%%%%%%%%%%%%%%%%%%%%%%%%%%%%%%%%%%%%%%%%%%%%%%%
\begin{abstract}
A universal scheme is introduced to speed up the dynamics of a driven open quantum system along a prescribed trajectory of interest. This framework generalizes counterdiabatic driving to open quantum processes. Shortcuts to adiabaticity designed in this fashion can be implemented in two alternative physical scenarios: one characterized by the presence of balanced gain and loss, the other involves non-Markovian dynamics with time-dependent Lindblad operators.  As an illustration, we engineer superadiabatic cooling, heating, and isothermal strokes for a two-level system, and provide a protocol for the fast thermalization of a quantum oscillator.
\blfootnote{$^*$These authors contributed equally to the work.}
\end{abstract}

%%%%%%%%%%%%%%%%%%%%%%%%%%%%%%%%%%%%%%%%%%%%%%%%%%%%%%%%%%%%%%%%

%\textit{Introduction.---}
Shortcuts to adiabaticity (STA) allow controlling the evolution of a quantum system without the requirement of slow driving \cite{Torrontegui13,delcampo19,guery-odelin2019}. The controlled speedup of quantum processes is broadly recognized as a necessity for the advance of quantum technologies, and STA have found a variety of applications,  including phase-space preserving cooling \cite{Chen10}, population transfer \cite{Demirplak03,Demirplak05},  and friction suppression in finite-time thermodynamics \cite{Deng13,delcampo14,Funo17}, to name some relevant examples.
To date, STA have been demonstrated in the laboratory using ultracold gases \cite{Schaff2010,Schaff2011a,Bason12,Rohringer15,Deng18pra,Deng18Sci,Diao18}, nitrogen-vacancy centers \cite{Zhang13,Koelbl19}, trapped ions \cite{An16}, superconducting qubits \cite{Wang18,Wang19}, and other systems \cite{Torrontegui13}.
 
Despite this remarkable progress, the use of STA has been predominantly restricted to tailor the dynamics of isolated driven systems. However, any physical system is embedded in a surrounding environment with which it can interact and exchange energy, particles, etc. In such a setting, the dynamics of the system is no longer-described by a Hamiltonian and is associated with a master equation \cite{BreuerBook}. A notable exception concerns the dynamics of an isolated system conditional to a given subspace of interest. The dynamics can then be described in terms of a non-Hermitian Hamiltonian,  that generates loss and gain when the system leaves the subspace of interest and returns to it, respectively \cite{PlenioKnight98}. Scenarios characterized by a balance of gain and loss arise naturally, e.g., in the presence of a non-Hermitian potential that breaks time-reversal symmetry but preserves parity-time-reversal symmetry, i.e., in $\mathpzc{PT}$-symmetric quantum mechanics \cite{Bender98,Ruter2010,Regensburger2012,Feng2012,Peng2014,Zhen2015}.

Recent efforts on developing STA in open quantum systems have predominantly  focused on mitigating decoherence  \cite{Torrontegui13,guery-odelin2019}. 
Perturbative methods have been put forward to inhibit unwanted transitions in two- and three-level systems \cite{ruschhaupt2012,kiely2014}, while the use of decoherence-free subspaces in open quantum systems allow one to mitigate decoherence \cite{Lidar98,wu2017,levy2018}, 

However, 
the use of STA to speed up open quantum processes  is expected to make possible a wide range  of applications such as design of novel cooling techniques, information erasure \cite{Boyd18}, or the engineering of superadiabatic quantum machines \cite{delcampo18}. 
In this context, the engineering of STA in systems described by non-Hermitian Hamiltonians has been advanced in Refs. \cite{ibanez2011,li2017,chen2018, impens2019} while the control by STA  of arbitrary nonunitary dynamics requires further progress.
A pioneering effort to this end introduced  fast control protocols for Markovian processes \cite{Vacanti14}. 
This guarantees an independent evolution for the different Jordan blocks forming the Lindblad operator, thus fulfilling the notion of adiabaticity for open system introduced in Ref. \cite{sarandy2005}. 
More recently,  the fast thermalization of a harmonic oscillator has been proposed via 
the reverse engineering of a non-adiabatic Markovian master equation \cite{Dann19} and engineered dephasing \cite{dupays20}. 
A related study has shown the possibility of speeding up the thermalization of  a system oscillator locally coupled to a harmonic bath \cite{Villazon19}.  Engineering of the system-bath coupling has also been proposed to speed-up isothermal processes \cite{pancotti2019}. The fast driving between equilibrium and squeezed states has also been presented \cite{dupays2020dynamical}.

In this paper, we introduce a universal scheme to engineer STA in arbitrary open quantum systems. Our work provides a generalization of the counterdiabatic driving technique \cite{Demirplak03,Demirplak05,Berry09} to open quantum processes.   To this end, we consider the evolution of a quantum system  described by a mixed state along a prescribed trajectory of interest. We then find the equation of motion  that generates the desired dynamics. The latter can be recast in terms of the nonlinear evolution of a system in the presence of balanced gain and loss. Alternatively, the dynamics can be associated with a non-Markovian master equation with time-dependent Lindblad operators whose explicit form is determined by the prescribed trajectory. We demonstrated this framework by discussing the controlled open quantum dynamics of a two-level system and a driven quantum oscillator.

%%%%%%%%%%%%%%%%%%%%%%%%%%%%%%%%%%%%%%%%%%%%%%%%%%%%%%%%%%%%%%%%
\section{STA by counterdiabatic driving}
Consider a quantum evolution of interest described by the mixed state
\begin{equation}
\label{varrhot}
\varrho(t)=\sum_{n=1}^{r}\lambda_n(t)|n_t\ra\la n_t|,
\end{equation}
of finite rank $r={\rm rank}(\varrho)$. 
%denotes the rank of the density matrix. 
We pose the problem of enforcing the evolution of the system through this trajectory.

Under unitary dynamics, eigenvalues of the density matrix remain constant, $\lambda_n(t)=\lambda_n(0)$---denoted briefly as $\lambda_n$. The equation of motion for the density matrix in this case reads
\begin{equation}
 \partial_{t}\varrho(t)=\sum_n\lambda_n \left(|\partial_tn_t\ra\la n_t|+|n_t\ra\la\partial_t n_t|\right),
\end{equation}
and can be recast as a Liouville-von Neumann equation, $\partial_{t}\varrho(t)=-i[{H}_1(t),\varrho(t)]$ (with $\hbar=1$), whenever the dynamics is generated by the Hamiltonian
\begin{equation}
{H}_{\rm 1}(t)= i \sum_n\left( |\partial_t n_t\ra\la n_t|{-}\la n_t|\partial_t n_t\ra |n_t\ra\la n_t|\right).
\end{equation}
This Hamiltonian generates parallel transport along each of the eigenstates $|n_t\ra$ and is often used in  proofs of the adiabatic theorem \cite{Kato50,Avron87}.

In the context of control theory, the derivation of ${H}_{\rm 1}(t)$ can be systematically achieved by the so-called counterdiabatic (CD) driving technique,  also known as transitionless quantum driving~\cite{Demirplak03,Demirplak05,Berry09}. Specifically, CD assumes that $|n_t\ra$ are the eigenstates of a reference system ${H}_0(t)$ that can be controlled by the auxiliary field ${H}_1(t)$ so that the full dynamics is actually generated by ${H}_0(t)+{H}_1(t)$. 
Yet, in the most general setting, the instantaneous eigenstates used in the specification of the trajectory (\ref{varrhot}) need not be the eigenstates of 
the physical Hamiltonian of the system. To identify a reference Hamiltonian in this case, we choose $\varrho(t)$ to evolve as a thermal state,
\beqa
\varrho(t)= e^{-\beta {H}_0(t)} / Z_0(t),
\eeqa
where $Z_0(t)=\tr[e^{-\beta {H}_0(t)}]$ denotes the partition function, and $\beta$ is the inverse temperature (assuming $k_{\textsc{b}}=1$). With this definition, the spectral decomposition of the reference Hamiltonian reads
\beqa
& {H}_0(t)=\sum_n E_n |n_t\ra\la n_t|,
\eeqa
where the eigenvalues $E_n=-\beta^{-1}\log(Z_0 \lambda_n)$ are time-independent, and so is the partition function. By construction $[{H}_0(t),\varrho(t)]=0$, and the state $\varrho(t)$ is a solution of
\beqa
 \partial_{t}\varrho(t)=-i[{H}_{\textsc{cd}}(t),\varrho(t)],
\eeqa
where ${H}_{\textsc{cd}}(t)={H}_0(t)+{H}_1(t)$.

%%%%%%%%%%%%%%%%%%%%%%%%%%%%%%%%%%%%%%%%%%%%%%%%%%%%%%%%%%%%%%%%
\section{CD driving of open quantum systems \label{sec:2}}
In what follows we shall focus on the case where the eigenvalues of the density matrix are time-dependent. The von Neumann entropy of the state is then a function of time, and the dynamics is generally open and nonunitary. Indeed, for an arbitrary change of the eigenvalues $\{\lambda_n\}$ the dynamics is generally non-trace-preserving.

For a given time-dependence of $\lambda_n(t)$, the equation of motion for the trajectory $\varrho(t)$ can be analogously derived as
\begin{equation}
\label{geom}
\partial_{t}\varrho(t){=}-i[{H}_{\textsc{cd}}(t),\varrho(t)]{+} \textstyle{\sum_n}\partial_{t}\lambda_n(t) |n_t\ra\la n_t|.
\end{equation}
The dynamics is trace-preserving whenever $\sum_n\lambda_n(t)=1$, i.e., $\textstyle{\sum_n}\partial_{t}\lambda_n(t)=0$.
The equation of motion (\ref{geom}) admits several physical interpretations that we discuss below.

\subsection{Mixed evolution under balanced gain and loss}
The additional term in Eq. (\ref{geom}) can be associated with the anti-Hermitian operator 
\begin{equation}
-i{\Gamma}(t)=\frac{i }{2}\sum_n \frac{\partial_{t}\lambda_n(t)}{\lambda_n(t)} |n_t\ra\la n_t|.
\end{equation}
The equation of motion for $\varrho(t)$ is then generated by the full non-Hermitian Hamiltonian ${H}(t)={H}_{\textsc{cd}}(t)-i{\Gamma}(t)$, i.e.,
\beqa
\label{gain-loss}
\partial_{t}\varrho(t)&=&-i\big(H(t)\varrho (t)-\varrho (t) H^{\dag}(t)\big)\nonumber\\
&=&-i\big[H_{\textsc{cd}}(t),\varrho (t)\big]-\big\{\Gamma(t),\varrho (t)\big\}.
\label{gain-loss-2}
\eeqa
For arbitrary $\{\lambda_n\}$,  this evolution is not necessarily norm-preserving and the norm varies at a rate
\begin{equation}
\partial_{t}\tr[\varrho(t)]=-2\tr[{\Gamma}(t)\varrho(t)]=\textstyle{\sum_n} \partial_{t}\lambda_n(t).
\end{equation}

A norm-preserving evolution through the trajectory $\varrho(t)$ is governed by the modified equation of motion
\beqa
\label{bglme}
\partial_{t}\varrho&=&-i({H}\varrho -\varrho {H}^\dag)- \partial_{t}\tr[\varrho] \,\varrho\nonumber \\
&=&-i\big[{H}_{\textsc{cd}},\varrho \big]+\big(2\la\Gamma\ra\varrho -\big\{{\Gamma},\varrho\big\}\big),
\eeqa
where $\langle \Gamma\rangle =\tr[\Gamma \varrho]$ and the time-dependence of all terms has been dropped for brevity. Note that the resulting equation is nonlinear in the quantum state $\varrho$. This dynamics thus takes the form of a mixed-state evolution in the presence of balanced gain and loss \cite{Brody12} with a time-dependent generator \cite{Gong13}. 
Balanced gain and loss arises naturally in the study of $\mathpzc{PT}$-symmetric quantum systems \cite{Bender98}, that can be used to describe  a  variety of experiments \cite{Ruter2010,Regensburger2012,Feng2012,Peng2014,Zhen2015}.
  
%%%%%%%%%%%%%%%%%%%%%%%%%%%%%%%%%%%%%%%%%%%%%%%%%%%%%%%%%%%%%%%%
  \subsection{Lindblad-like form}Considering the prescribed trajectory \eqref{varrhot} and its derivative \eqref{geom}, one can recast the incoherent part 
\begin{equation}
\mathpzc{D}_{\textsc{cd}}(\varrho)=  \textstyle{\sum_n} \partial_{t}\lambda_n(t) |n_t\ra\la n_t|
\end{equation}
as an auxiliary CD dissipator in a Lindblad-like form for a trace-preserving trajectory. Assuming a trace-preserving evolution, $\sum_n \partial_t \lambda_n(t) =0$, we find the time-dependent Lindblad operators and rates as (see the appendixes)
 \begin{subequations}
 \begin{align}
{L}_{mn}(t)&=\ket{m_t}\bra{n_t},\\
\gamma_{mn}(t)&=\frac{\partial_{t}\lambda_m(t)}{r\lambda_n(t)},
\label{rates}
 \end{align}
 \end{subequations}
that are determined by (the spectral resolution of) $\varrho(t)$---and thus state-dependent. The resulting master equation 
\begin{align}
\label{nmme}
\partial_{t}\varrho=&-i[H_{\textsc{cd}},\varrho] \\
&+\textstyle{\sum_{mn}} \gamma_{mn}\big({L}_{mn}\varrho {L}_{mn}^\dag-\frac{1}{2}\{{L}_{mn}^\dag {L}_{mn},\varrho\}\big) \nonumber
\end{align}
is generally non-Markovian, because of possibly negative rates. We remark that the existence of a Lindblad-like master equation for an arbitrary dynamics has recently been proven in Ref. \cite{ULL}.  However,  in the representation (\ref{geom}) like a Lindblad-like master equation, the anticommutator term in Eq. (\ref{nmme}) identically vanishes and the dissipator reduces exclusively to jumps in the instantaneous eigenbasis.

The equivalence of Eqs. (\ref{bglme}) and (\ref{nmme}) shows that the nonlinear evolution of a mixed state under balanced gain and loss can be represented by a nonlinear and generally non-Markovian master equation with time-dependent Lindblad operators, determined by choice of the trajectory \eqref{varrhot}.

We note that the time-evolution operator generated by the CD Hamiltonian takes the form \cite{Berry09}
\beqa
{U}_{\textsc{cd}}(t,0)=\textstyle{\sum_n} e^{i\phi_n(t)}|n_t\ra\la n_0|,
\eeqa
where the time-dependent phase $\phi_n(t)$ is the sum of the dynamical and geometric contributions.
In the co-moving frame associated to ${U}_{\textsc{cd}}(t,0)$, the master equation for $\tilde{\varrho}(t)={U}_{\textsc{cd}}^{\dag}(t,0)\varrho (t) {U}_{\textsc{cd}}(t,0)$ takes the simple form
\begin{equation}
\partial_{t}\tilde{\varrho}\,{=} \sum_{mn} \gamma_{mn}\big({\tilde{L}}_{mn}\tilde{\varrho} {\tilde{L}}_{mn}^\dag{-}\frac{1}{2}\{{\tilde{L}}_{mn}^\dag {\tilde{L}}_{mn},\tilde{\varrho}\}\big),
\end{equation}
with ${\tilde{L}}_{mn}=\ket{m_0}\bra{n_0}$. As a result, the time-dependent Lindblad operators $\{{L}_{mn}\}$ map to the time-independent ones $\{{\tilde{L}}_{mn}\}$, while keeping the same rates $\gamma_{mn}(t)$. This feature is specific to the superadiabatic driving of open quantum systems  and differs from the general case that leads to more complex time-dependent Lindblad operators \cite{BreuerBook}.

%%%%%%%%%%%%%%%%%%%%%%%%%%%%%%%%%%%%%%%%%%%%%%%%%%%%%%%%%%%%%%%%
\section{Quantum speed limit for STA in open quantum processes}

Time-energy uncertainty relations identify characteristic time scales in a physical process. 
Speed limits  sharpen this identification by providing a minimum time for a physical processes to occur in terms of the generator of the evolution. We next show how speed limits  relate the operation time of a protocol to the amplitude of the required unitary and nonunitary CD terms. The geometric formulation of the quantum speed limit \cite{Funo:QSL-OpenSystem}  states that 
\beqa
\label{QSLFuno}
\tau \geqslant D(\varrho(0),\varrho(\tau)) / \langle{\sqrt{g_{tt}}\rangle}, 
\eeqa
where $g_{tt}$ is the metric for a given distance $D$, and the time average $\langle{\sqrt{g_{tt}}\rangle}_{\tau}=(1/\tau)\int_0^{\tau}dt\,\sqrt{g_{tt}}$ upper bounds the speed of evolution. 

The quantum Fisher information $\mathpzc{F}$ is the metric (with a $1/4$ prefactor) associated with the Bures distance between quantum states 
\beqa
D_B(\varrho_1,\varrho_2)=\left[2(1-F(\varrho_1,\varrho_2)\right]^{1/2},
\eeqa 
that is defined in terms of the fidelity $F(\varrho_1,\varrho_2)=\mathrm{Tr}\sqrt{\sqrt{\varrho_1}\varrho_2\sqrt{\varrho_1}}$ between $\varrho_1$ and $\varrho_2$ \cite{Rezakhani-Abasto}.
The speed limit (\ref{QSLFuno}) implies that the driving time of the process is constrained by the ratio of the distance between quantum states $D(\varrho(0),\varrho(\tau))$ and the velocity at which is traversed $\langle{\sqrt{g_{tt}}\rangle}$.

From Eq.~\eqref{gain-loss}, we can identify $-2i{H}$ as a non-Hermitian symmetric logarithmic derivative, satisfying $2\partial_{t}\varrho=\mathbbmss{L}\varrho+\varrho\mathbbmss{L}^\dag$  \cite{Alipour-convexity}, based on which an upper bound on the quantum Fisher information is obtained as $\mathpzc{F}=\mathrm{Tr}[\varrho\mathbbmss{L}^{2}] \leqslant 4\,\mathrm{Tr}[{H}\varrho {H}^{\dag}]$. As a result, the quantum speed limit reads
\begin{align}
\tau \geqslant\frac{D_{B}(\varrho(0),\varrho(\tau))}{4\langle \mathrm{Tr}[(H_{\textsc{cd}}-i \Gamma)\varrho(t) (H_{\textsc{cd}}+i \Gamma^{\dag})]^{1/2}\rangle_{\tau}}.
\end{align}
The minimum time to implement a STA driving the system  from $\varrho(0)$ to $\varrho(\tau)$ is thus not only governed by the Hermitian system Hamiltonian $H_{\textsc{cd}}$, but as well by the term $\Gamma$ governing gain and loss.

Alternatively, using the trace distance rather than the Bures distance, the relevant metric is $g_{tt}=\|\partial_{t}\varrho \|_1^2\equiv (\mathrm{Tr}[\sqrt{(\partial_{t}\varrho) ^2}])^2$. Using Eqs. \eqref{gain-loss} and \eqref{nmme} for $\partial_{t}\varrho $ and the triangle inequality, one obtains $\|\partial_{t}\varrho \| \leqslant \|[{H}_{\textsc{cd}},\varrho]\|+\|\{{\Gamma},\varrho \} \|$ for the gain-loss equation and $\|\partial_{t}\varrho \| \leqslant \|[{H}_{\textsc{cd}},\varrho]\|+\|\mathpzc{D}_{\textsc{cd}}\|$ for the Lindblad-like equation. In all of these bounds, both the CD Hamiltonian and dissipator set the speed of evolution.

%%%%%%%%%%%%%%%%%%%%%%%%%%%%%%%%%%%%%%%%%%%%%%%%%%%%%%%%%%%%%%%%
\section{Examples}
\subsection{Strokes for a two-level system}
Consider a two-level system described by a time-dependent Hamiltonian 
\beqa
{H}_0(t)=\frac{1}{2}\big(\Delta(t){\sigma}_z+\Om(t){\sigma}_x\big),
\eeqa
where ${\sigma}_z$ and ${\sigma}_x$ are the Pauli matrices. The instantaneous eigenstates read $E_\pm(t)=\pm\sqrt{\Om^2(t)+\Delta^2(t)}/2=\pm |\Om(t)|/(2\sin\theta(t))$, where $\theta(t)={\rm arctan}(\Om(t)/\Delta(t))$ and  the corresponding eigenstates are
\beqa
|+_t\ra&=&\cos(\theta(t)/2)|0\ra+\sin(\theta(t)/2)|1\ra,\nonumber\\
|-_t\ra&=&\sin(\theta(t)/2)|0\ra-\cos(\theta(t)/2)|1\ra,
\eeqa
with $\sigma_z |0\rangle=|0\rangle$ and $\sigma_z |1\rangle=-|1\rangle$. We consider the system to be described by the time-dependent mixed state $\varrho(t)=\sum_{\alpha=\pm}\lambda_{\alpha}(t)|\alpha_t\ra\la \alpha_t|$. Thus, the target trajectory $\varrho$ is already diagonal in the eigenbasis of the uncontrolled system Hamiltonian ${H}_0(t)$. The auxiliary control term required to guide the dynamics is known to be of the form \cite{Demirplak03,Demirplak05,Berry09}
\beqa
\label{H1TLS}
{H}_1=\frac{1}{2}\frac{\partial_{t}\Delta\,\Om-\partial_{t}\Om\,\Delta}{\Om^2+\Delta^2}\sigma_y,
\eeqa
so that the full dynamics is generated by ${H}_{\textsc{cd}}={H}_0+{H}_1$. The dynamics is open when the eigenvalues $\lambda_{\pm}$ are time-dependent. 

The first approach we have introduced relies on the presence of gain and loss, for which the dynamics is generally no longer trace-preserving, i.e., $\lambda_{_-}+\lambda_{_+}$ is time-dependent and different from unity. Such evolution  is generated by the non-Hermitian Hamiltonian ${H}={H}_{\textsc{cd}}-i{\Gamma}$, where
\beqa
\Gamma=\frac{\partial_{t}\lambda_{_+}}{2}\Big(\frac{1}{\lambda_{_-}}|-_t\ra\la-_t|-\frac{1}{\lambda_{_+}}|+_t\ra\la+_t|\Big).
\label{28}
\eeqa
Under balanced gain and loss, the trace-preserving property is restored by the nonlinear equation (\ref{bglme}) with this choice of ${\Gamma}$. 

Alternatively, STA in an open two-level system can be associated with a Lindblad-like master equation with the Lindblad operators
\beqa
{L}_{+-}(t)=|+_t\ra\la -_t|, \quad
{L}_{-+}(t)={L}^{\dag}_{+-}(t).
\eeqa
The rates  are given by $\gamma_{+-}(t)=\frac{\partial_{t}\lambda_{_+}}{2\lambda_{-}}$ and $\gamma_{-+}(t)=\frac{\partial_{t}\lambda_{_-}}{2\lambda_{+}}$.

Assume that the system is initially prepared in a thermal state at inverse temperature $\beta(0)$, $\varrho(0)=\sum_{\alpha=\pm}\lambda_\alpha(0)|\alpha_0\ra\la \alpha_0|$, where $\lambda_\alpha=e^{-\beta(0) E_\alpha(0)}/Z(0)$ with $Z(0)=e^{-\beta(0) E_{-}(0)}+e^{-\beta(0) E_+(0)}$.
We focus on description of thermodynamic protocols for which the target trajectory $\varrho(t)$ is an instantaneous thermal state with inverse temperature $\beta(t)$, i.e.,
\begin{equation}
\varrho(t)=\sum_{\alpha=\pm} \frac{e^{\alpha\frac{\beta(t)}{2}\sqrt{\Om^2(t)+\Delta^2(t)}}}{2\cosh[\frac{\beta(t)}{2}\sqrt{\Om^2(t)+\Delta^2(t)}]}|\alpha_t\ra\la \alpha_t|.
\end{equation}

One can engineer different processes of interest which are of this type. For example, in a superadiabatic isothermal stroke, the state is always in a thermal form at a given reference inverse temperature $\beta(t)=\beta(0)$, regardless of the rate at which ${H}_0(t)$ is driven. Nonadiabatic excitations are cancelled by the auxiliary term ${H}_1$ in Eq. (\ref{H1TLS}), while the thermal form of $\lambda_{\pm}(t)$ is guaranteed by the Lindblad operators and rates.
For arbitrary $\Delta(t)$ and $\Omega(t)$, they read
\begin{eqnarray}
&&{L}_{+-}=|+\ra\la -|,\quad {L}_{-+}=|-\ra\la +|, \label{strokes}\\
&&\gamma_{\alpha\alpha'}=\frac{\alpha~\beta}{2} \frac{\Delta\,\partial_{t}\Delta+\Omega\,\partial_{t}\Omega }{\sqrt{\Delta
   ^2+\Omega ^2}} \big(e^{\alpha'\beta  \sqrt{\Delta^2+\Omega^2}}+1\big)^{-1}, \nonumber
\end{eqnarray}
where $\alpha, \alpha' \in \{\pm\}$ and $\alpha\neq \alpha'$.
%****************Exp Proposal******************************
%%%%%%%%%%%%%%%%%%%%%%%%%%%%%%%%%%%%%%%%%%%%%%%%%%%%%%%%%%%%%%%%
\begin{figure}[t]
\includegraphics[width=\linewidth]{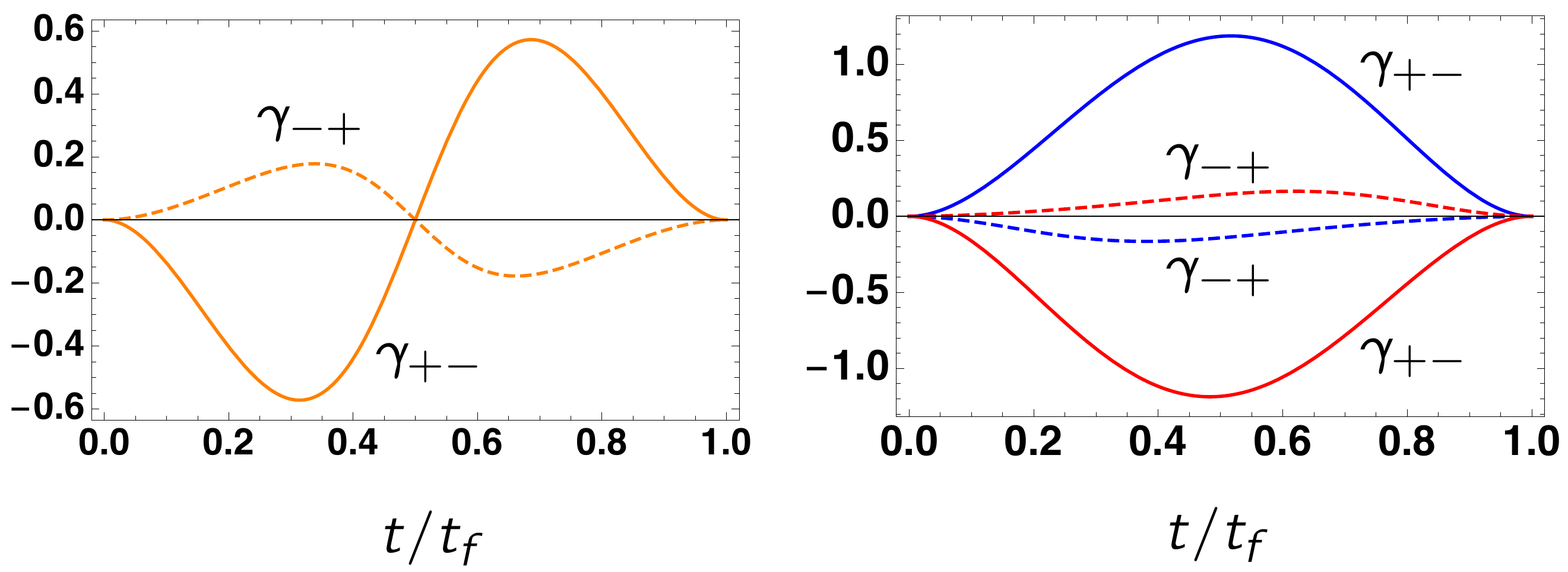}
\caption{Left: Time-dependence of the rates in an isothermal process at inverse temperature $\beta=1$, keeping $\Om$ constant with initial $\Delta(0)=1$ and final $\Delta(t_f)=-1$. Right: Time-dependence of the rates for the superadiabatic cooling (blue) and heating (red) of a two-level system. Taking $\Delta=\Omega$ as the unit of frequency, the process corresponds to cooling a thermal state from $\beta(0)=1$ to $\beta(t_f)=2$ and heating a thermal state from $\beta(0)=2$ to $\beta(t_f)=1$.}
\label{FigRates}
\end{figure}
%%%%%%%%%%%%%%%%%%%%%%%%%%%%%%%%%%%%%%%%%%%%%%%%%%%%%%%%%%%%%%%%
%------------------------------------------------

A typical modulation in time is shown in Fig. \ref{FigRates} for a two-level system to evolve along STA for an isothermal stroke induced by driving of $\Delta(t)$ while keeping $\Om$ constant. Specifically, $\Delta(t)$ is chosen as a fifth-order polynomial in time interpolating between the initial and final values. The rates have opposite signs, vanish identically at the avoided crossing, and flip signs during the subsequent evolution.

It is possible to look as well for  cooling  and heating  protocols characterized by a time-dependent inverse temperature $\beta(t)$ keeping ${H}_0$ constant, as required, e.g., in a quantum Otto cycle. In such a case, ${H}_1$ vanishes, and the cooling and heating strokes are implemented by time-independent Lindblad operators with time-dependent rates,
\beqa
\gamma_{\alpha \alpha'}(t)&=&\frac{\alpha~\partial_{t}\beta (t)}{2}\frac{\sqrt{\Delta ^2+\Omega ^2} 
}{e^{-\alpha' \sqrt{\Delta ^2+\Omega ^2} \beta (t)}+1},
\eeqa
where $\alpha, \alpha' \in \{\pm\}$~and $\alpha\neq \alpha'$. 
The time-dependence of the rates is explicitly illustrated for both cooling and heating processes in Fig. \ref{FigRates}, for constant values of $\Delta$ and $\Omega$, and $\beta(t)$ interpolating between $\beta(0)$ and $\beta(t_f)$ again as a fifth-order polynomial. The non-Markovian character of the evolution is manifest given the time-dependence of the Lindblad operators and the opposite sign of the corresponding rates.

Beyond these two prominent examples, more general strokes can be considered.  The required Lindblad operators in the most general setting are provided in the appendixes. We also note that in all cases the corresponding operator $\Gamma$ associated with gain and loss can be conveniently expressed  in terms of the rates as
\begin{equation}
{\Gamma}(t)=\gamma_{-+}(t)|+_t\ra\la+_t|-\gamma_{+-}(t)|-_t\ra\la-_t|.
\end{equation}

In the following, we consider another example in which the real physical dynamics of the system keeps its state always in the Gibbsian form with a time-dependent temperature. 

%%%%%%%%%%%%%%%%%%%%%%%%%%%%%%%%%%%%%%%%%%%%%%%%%%%%%%%%%%%%%%%%
\subsection{STA for equilibration of a thermalizing atom}
Consider a two-level atom in a thermal bosonic bath at inverse temperature $\beta_B(0)$.
The dynamics of the atom under some conditions can be described by \cite{SciRep:Alipour-corr, Rajagopal, Book:Carmichael}
\begin{align}
\label{lindblad-thermal-2}
&\partial_{t}\varrho _{S}= -i\big[H_S,\varrho_{S}\big]\\
&+\textstyle{\sum_{j,k|(j\neq k)}} \gamma_{jk} (L_{jk}\varrho_S L_{jk}^{\dag}-\frac{1}{2}\{L_{jk}^{\dag}L_{jk},\varrho_S\}), \nonumber
\end{align}
where $j,k \in \{0,1\}$, and 
\begin{align}
&\hskip-1.5mm H_{S}=\frac{\omega_0}{2}\sigma_{z},~L_{10}=|1\rangle \langle 0|=\sigma_-,~ L_{01}=|0\rangle \langle 1|=\sigma_+, \label{DJayCum-Lind-rates}\\
&\hskip-1.5mm \gamma_{10}=\gamma \big(\bar{n}\left(\omega_0,\beta_B(0)\right) + 1\big),~~~\gamma_{01}=\gamma \bar{n}. \nonumber
\end{align}
Here, $\bar{n}\big(\omega_0,\beta_B(0)\big) = (\mathrm{e}^{\beta_B\omega_0}-1)^{-1}$
is the mean boson number in a mode with frequency $\omega_{0}$, and $\gamma$ is a time-independent constant indicating the strength of the coupling between the atom and the thermal bath. 

%%%%%%%%%%%%%%%%%%%%%%%%%%%%%%%%%%%%%%%%%%%%%%%%%%%%%%%%%%%%%%%%
\begin{figure}[tp]
\includegraphics[width=\linewidth]{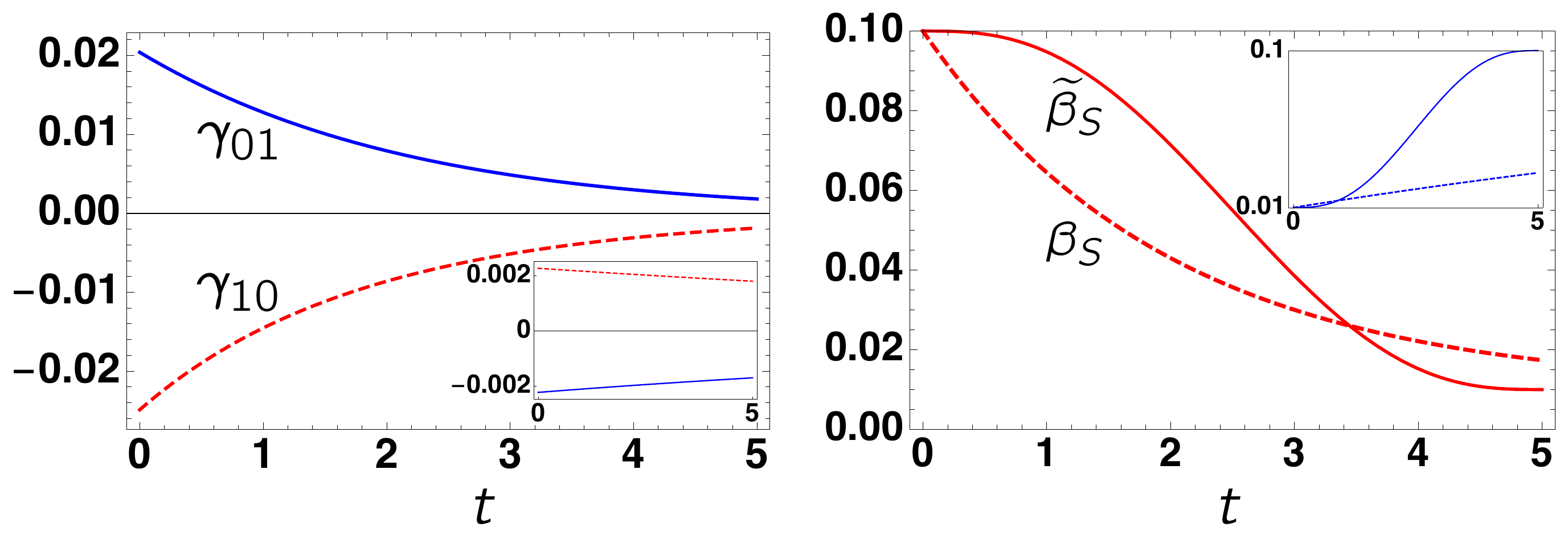}
\caption{Left: Time dependence of the rates  $\gamma_{10}$ (red, dashed) and $\gamma_{01}$ (blue, solid) in Eq. \eqref{nmme} for a thermalizing atom in the case of heating, when $\beta_S = 0.1$, $\beta_B = 0.01$, $\omega_0$ = 2, and $\gamma= 0.005$. The inset corresponds to the cooling case, with $\beta_S \leftrightarrow \beta_{B}$. With these parameters, the rates for the Lindblad master equation of the subsystem in Eqs.~\eqref{lindblad-thermal-2} and \eqref{DJayCum-Lind-rates} become constant; $\gamma_{01}=4.95$ and $\gamma_{10}=5.05$. Right: Inverse temperature of the system for the Lindblad master equation (dashed) and for the STA (solid). The dependence in time is shown for the case of heating (see inset for the cooling case) with the same set of parameters as in the left panel.}
\label{fig:rates}
\end{figure}
%%%%%%%%%%%%%%%%%%%%%%%%%%%%%%%%%%%%%%%%%%%%%%%%%%%%%%%%%%%%%%%%

If the atom is initially in a thermal state $\varrho_S(0)=e^{-\beta_S(0) H_S}/Z_S(0)$, its instantaneous state is obtained by solving the above master equation, which gives a Gibbsian thermal state $\varrho_{S}(t)=e^{-\beta_{S}(t)H_{S}}/Z_{S}(t)$, 
with
\begin{equation}
\label{beta}
\beta_S(t){=}\frac{-1}{\omega_0} \log \frac{1{-}\mathrm{e}^{- \tilde{\gamma}t} \mathrm{tanh}\Theta_S{+}(\mathrm{e}^{- \tilde{\gamma}t} {-}1)\mathrm{tanh}\Theta_B}{1{+}\mathrm{e}^{- \tilde{\gamma}t}\mathrm{tanh}\Theta_S{-}(\mathrm{e}^{- \tilde{\gamma}t} {-}1) \mathrm{tanh}\Theta_B}.
\end{equation}
Here $\Theta_k=\omega_0\,\beta_k(0)/2$ ($k\in\{S,B\}$), $\tilde{\gamma}= \gamma \coth\Theta_B$, and $Z_S(t)=\mathrm{Tr}[e^{-\beta_S(t)H_S}]$. 

Equation \eqref{nmme} suggests another dynamical equation realizing the same trajectory $\varrho_S(t)$. Since $H_0=H_S$ is time-independent, $H_1$ will be zero as well. The Lindblad operators are given in terms of the eigenstates of $H_S$ as $L_{mn}=|m\rangle\langle n|$ where $m,n \in \{0,1\}$, and the rates are obtained from Eq.~\eqref{rates} by considering that $\lambda_0=e^{-\beta_S(t) \omega_0/2}/Z_S(t)$ and $\lambda_1=e^{\beta_S(t)\omega_0/2}/Z_S(t)$ can also be identified simply as $\varrho_{00}$ and $\varrho_{11}$, respectively (see the appendixes). While the Lindblad operators here are equal to those in Eq.~\eqref{lindblad-thermal-2}, the rates in the Markovian master equation \eqref{lindblad-thermal-2} are positive constants. By contrast, the rates in Eq.~\eqref{nmme} are time-dependent and negative for some time intervals, as illustrated in Fig. \ref{fig:rates}. Nonetheless, in both cases equilibration with the bath at temperature $\beta_B$ takes infinite time. 

A STA in finite time $t_f$ can be associated with a trajectory $\tilde{\varrho}_S(t) =e^{-\tilde{\beta}_S(t)H_S}/Z_S(t)$ and a modified inverse temperature $\tilde{\beta}_{S}(t)$ satisfying $\tilde{\beta}_{S}(t_f)=\beta_B$. Using Eq.~\eqref{nmme}, the Lindblad operators remain unchanged, as in Eq.~\eqref{DJayCum-Lind-rates}, whereas the rates are obtained from Eq.~\eqref{rates} as $\gamma_{01}=-(\omega_0/4)~\partial_t\tilde{\beta}_{S}(t)\,e^{-\tilde{\beta}_S(t)\,\omega_0}$ and $\gamma_{10}=(\omega_0/4)\partial_t\tilde{\beta}_{S}(t)\,e^{\tilde{\beta}_S(t)\,\omega_0}$. In Fig. \ref{fig:rates}, the right panel shows the temperature for a typical function as $\tilde{\beta}_S(t)$ such that at $t_f=5$ the system state thermalizes, i.e., $\tilde{\varrho}_S(t_f)=e^{-\beta_B H_S}/\mathrm{Tr}[e^{-\beta_B H_S}]$. 
%The required modulation of the rates can be implemented using classical noise as a resource \cite{Chenu17,Khurana19}. fore Hermitian Lindbaldians
%%%%%%%%%%%%%%%

\subsection{Fast thermalization of a quantum oscillator}
We next consider the fast thermalization of a quantum oscillator using the general scheme presented in Sec. \ref{sec:2}.
%relates to the engineered master equation used in Ref. \cite{dupays20}. 
This illustrate an application of the proposed scheme for an infinite rank density matrix, that can be  implemented with current technology. 
Alternative approaches for the fast thermalization of an oscillator have been recently presented in Ref. \cite{Dann19,dupays20}.

Consider the time-dependent Hamiltonian $H_0 = \frac{\hat{p}^2}{2m} + \frac{1}{2}m \omega_t^2 \hat{x}^2$ with instantaneous thermal state $\varrho(t)= e^{- \beta_t H_0}/ \Tr[e^{- \beta_t H_0}]$. 
In the basis of the instantaneous Fock states $\ket{n_t}$, 
 the thermal state is diagonal, $\varrho(t) = \sum_n \lambda_{n}(t) \ket{n_t} \bra{n_t}$,  with probabilities $\lambda_{n}(t) = u_t^n (1-u_t)^{-1}$ that are generally time-dependent due to the modulation of the  frequency and temperature, where $u_t = e^{- \beta_t \hbar \omega_t}$. 
The thermal state evolves according to Eq.~\eqref{geom},
%\begin{equation}\label{eq:MEsigma}
%\cancel{\partial_{t}\varrho(t) =  -\frac{i}{\hbar}[H_0+H_1, \varrho(t)] + \sum_n \dot{\lambda}_{n}(t) \ket{n_t}\bra{n_t},}
%\end{equation} 
where the commutator $[H_0,\varrho(t)]=0$ and the  counterdiabatic  Hamiltonian term $H_1$ is given by \cite{Muga2010,Jarzynski13,delcampo-13}
\beqa
H_1 = -\frac{\dot{\omega}_t}{4 \omega_t} \{\hat{x}, \hat{p}\}.
\eeqa
This term can in principle be engineered in a trapped ion as  suggested in Ref.~\cite{Funo17}.
We show below a scheme for implementing in the laboratory the unitarily equivalent trajectory $\tilde{\varrho}(t) = U_x \varrho(t)U_x^\dag$, where  
\beqa
U_x = e^{i \frac{m }{2 \hbar}\alpha_t\hat{x}^2},  
\eeqa
and $\alpha_t$ is a frequency to be determined. Such  trajectory maps an initial thermal state into a final thermal state of different temperature provided $\alpha_t$ vanishes at the beginning and end of the protocol.  Direct computation of its time derivative   gives %\begin{equation}\label{ME1}
$ \partial_{t}\tilde{\varrho} = \frac{i}{\hbar} \Big[\frac{m}{2}\dot{\alpha}_t~\hat{x}^2, \tilde{\varrho} \Big] + U_x(\partial_{t}\varrho) U_x^\dagger$, 
which admits a form similar to  Eq.~\eqref{geom}, i.e, 
 \begin{equation}\label{ME2}
 \partial_{t}\tilde{\varrho}= -\frac{i}{\hbar}[\tilde{H}_{\textsc{cd}}, \tilde{\varrho}] +  \tilde{\mathpzc{D}}_{\textsc{cd}}(\tilde{\varrho}),
\end{equation}
where the counterdiabatic Hamiltonian in the rotating frame reads
\begin{align} \label{HCD1}
\tilde{H}_{\textsc{cd}} &= i\hbar \dot{U}_x U_x^{\dag}+ U_x (H_0+ H_1)U_x^{\dagger}, 
%&\textcolor{red}{\equiv -\frac{m}{2}\dot{\alpha}  \hat{x}^2+ U_x (H_0+ H_1)U_x^{\dagger}}
\end{align}
and  the dissipator  is given by 
\begin{equation} \label{Dcd1}
 \tilde{\mathpzc{D}}_{\textsc{cd}}(\tilde{\varrho})= \textstyle{\sum_n} \dot{\lambda}_{n}(t) U_x \ket{n_t}\bra{n_t}U_x^{\dagger}.
\end{equation}

By explicit computation,  the counterdiabatic Hamiltonian Eq.~(\ref{HCD1}) can be recast as 
\begin{equation}
\tilde{H}_{\textsc{cd}}  = \frac{\hat{p}^2}{2m}  +\frac{1}{2} m \tilde{\omega}_t^2 \hat{x}^2 - \left(\frac{\alpha_t}{2} + \frac{\dot{\omega}_t}{4 \omega_t}\right) \{\hat{x},\hat{p}\},
\end{equation}
with  time-dependent frequency 
\begin{equation} \label{fqcontrol}
\tilde{\omega}_t^2 =\omega_t^2  + \alpha_t^2+  \alpha_t \frac{\dot{\omega}_t}{\omega_t} - \dot{\alpha}_t. 
\end{equation}
It proves convenient to define $\alpha_t=\Omega_t-\dot{\om}_t/(2\om_t)$, so that
\begin{equation}
\tilde{H}_{\textsc{cd}}  = \frac{\hat{p}^2}{2m}  +\frac{1}{2} m \tilde{\omega}_t^2 \hat{x}^2 - \frac{\Om_t}{2}  \{\hat{x},\hat{p}\}.
\end{equation}

As shown in App. \ref{appC}, by further choosing 
\begin{equation} \label{assumption}
\Omega_t = -\frac{1}{2}\frac{\dot{\omega}_t}{\omega_t} + \frac{\dot{u}_t}{1-u_t^2},
\end{equation}
 the dissipator  in the rotating frame equals
\beqa\label{Dcd_HO}
  \tilde{\mathpzc{D}}_{\textsc{cd}}(\tilde{\varrho})& =&\frac{1}{i\hbar}\left[\frac{\Om_t}{2}\{\hat{x},\hat{p}\}-m\alpha_t\Om_t\hat{x}^2,\tilde{\varrho}\right]\nonumber\\
  & & -\gamma_t[\hat{x},[\hat{x},\tilde{\varrho}]],
  \eeqa
with a time-dependent dephasing strength  
\beqa
\label{hogteq}
\gamma_t = \frac{m \omega_t}{\hbar}\frac{\dot{u}_t}{(1-u_t)^2}.
\eeqa 
Combining the explicit forms of $\tilde{H}_{\textsc{cd}}$ and $\tilde{\mathpzc{D}}_{\textsc{cd}}(\tilde{\varrho})$ in 
Eq.~\eqref{ME2} results in the master equation of a time-dependent quantum oscillator  subject to dephasing in the coordinate representation, i.e., 
\begin{eqnarray}
 \partial_{t}\tilde{\varrho}&=&\frac{1}{i\hbar}\left[\frac{\hat{p}^2}{2m} + \frac{1}{2}m \tilde{\omega}_{\textsc{cd}}^2 \hat{x}^2 , \tilde{\varrho}\right] - \gamma_t [\hat{x}, [\hat{x},\tilde{\varrho}]]. \label{ME3}\nonumber\\
\end{eqnarray}
where
\beqa
\label{omcd}
 \tilde{\omega}_{\textsc{cd}}^2&=&\om_t^2-\alpha_t^2-\dot{\alpha}_t\\
& =&  \left[\omega_t^2  -\frac{3}{4}\left(\frac{\dot{\omega}_t}{\omega_t}\right)^2+\frac{\ddot{\omega}_t}{2\omega_t}\right] -\Om_t^2-\dot{\Om}_t+\Om_t\frac{\dot{\omega}_t}{\omega_t},\nonumber
\eeqa
and $\gamma_t$ is given by Eq. (\ref{hogteq})
and   $\Omega_t$ by Eq. (\ref{assumption}).
The case of unitary dynamics in which the eigenvalues  $\{\lambda_n\}$ are constant corresponds to $\Om_t=0$, i.e., $\alpha_t=-\dot{\omega}_t/(2\omega_t)$. The first term in square brackets on the right-hand side (RHS) of Eq.~(\ref{omcd}) is indeed that used for the (local) counterdiabatic driving of a driven oscillator in the absence of coupling to a bath \cite{Ibanez12,delcampo-13,Funo17}.
The dynamics described by Eq.~(\ref{ME3}) generalizes the case of unitary evolution to account for the controlled driving  of an open quantum oscillator (i.e., when  the eigenvalues  $\{\lambda_n\}$ of the density matrix are time-dependent)  from an initial thermal state to a final thermal state in arbitrary time.
The implementation of a STA by counterdiabatic driving for the fast thermalization of a quantum oscillator is  achieved by a simultaneous modulation of the driving frequency and the dephasing strength.
The dynamics associated with Eq.~(\ref{ME3}) can be readily implemented in the laboratory. It requires the control of the trap frequency and dephasing strength. The latter can be engineered for $\gamma_t>0$ using noise as a resource via stochastic parametric driving, or through continuous quantum measurements, as recently proposed in Ref. \cite{dupays20}. While the counterdiabatic driving protocol derived here requires similar experimental resources to the ones for STA based on reverse-engineering of the dynamics \cite{dupays20}, the time modulations of the driving frequency and the dephasing strength need not be equal, and generally differ, between the two protocols. In addition, their experimental implementation is at reach with current technology in trapped ions \cite{Funo17,Smith18} and ultracold gases \cite{Schaff2010}.

%%%%%%%%%%%%%%%%%%%%%%%%%%%%%%%%%%%%%%%%%%%%%%%%%%%%%%%%%%%%%%%%
\begin{figure*}[tp]
\begin{center}
\includegraphics[width=0.8\linewidth]{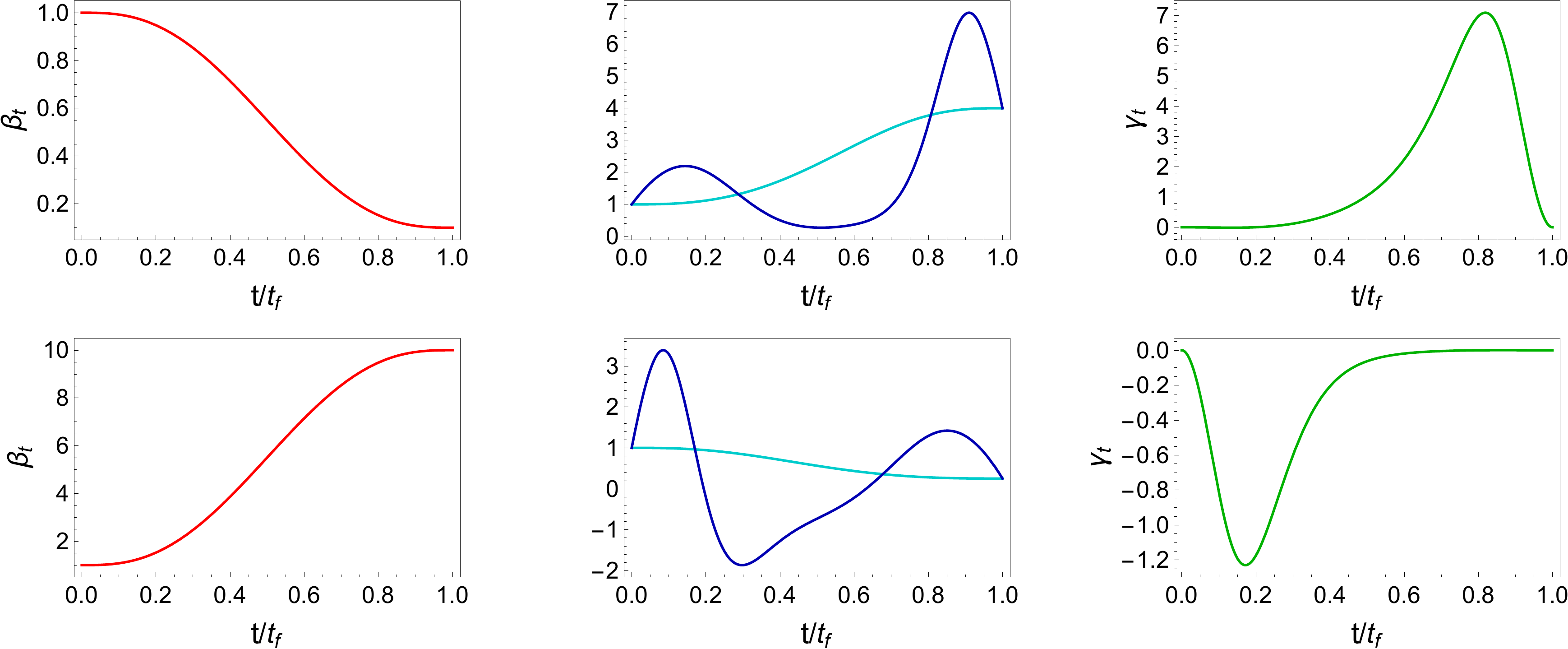}
\caption{Counterdiabatic driving of an open driven quantum  quantum oscillator. Heating stroke (top):  Left: Modulation of the inverse-temperature with $\beta_f=\beta_0/10$.  Center:  Monotonic reference modulation of the trapping square frequency $\om_t^2/\om_0^2$  compared with the nonmonotonic counterdiabatic modulation $ \tilde{\omega}_{\textsc{cd}}^2/\om_0^2$, with $\om_f=2\om_0$. Right: Time-dependent dephasing strength in units of $m\om_0/\hbar$. The bottom row shows the engineering of a cooling stroke with $\om_f=\om_0/2$ and $\beta_f=10\beta_0$. Negative values of $ \tilde{\omega}_{\textsc{cd}}^2/\om_0^2$ are associated with trap inversion.}
\label{figTDHO}
\end{center}
\end{figure*}
%%%%%%%%%%%%%%%%%%%%%%%%%%%%%%%%%%%%%%%%%%%%%%%%%%%%%%%%%%%%%%%%

To illustrate a specific protocol we consider a reference trajectory $\varrho(t)$ describing the evolution from an initial thermal state of frequency $\om_0$ at inverse temperature $\beta_0$ 
to a final thermal state with frequency $\om(t_f)=\om_f$ and  inverse temperature $\beta(t_f)=\beta_f$. 
For instance, $\varrho(t)$  can be specified by choosing the  interpolating ansatze
\beqa
\om_t&=&\om_0+ (\om_f-\om_0)[10s^3-15s^4+6s^5],\\
\beta_t&=&\beta_0+ (\beta_f-\beta_0)[10s^3-15s^4+6s^5],
\eeqa
with $s=t/t_f$, where $t_f$ is the duration of the process. The polynomial functions are monotonic as a function of time.
The required experimental controls to implement the unitarily equivalent trajectory $\tilde{\varrho}(t)$  are $ \tilde{\omega}_{\textsc{cd}}$ in Eq.~(\ref{omcd}) and $\gamma_t$ in Eq.~(\ref{hogteq}) with $u_t=e^{-\beta_t\hbar\om_t}$, shown in Fig. \ref{figTDHO}. Specifically, a heating stroke involving a trap compression shows that the required  counterdiabatic modulation of the trapping frequency exhibits a nonmonotonic behavior involving of sequence of  tight compressions and decompressions, overshooting the reference modulation.  Along the process, the dephasing strength takes predominantly positive values, thus suppressing coherences in the position eigenbasis.
counterdiabatic cooling strokes are more challenging to implement  than counterdiabatic heating strokes. 
First, the dephasing strength takes negative values  throughout the cooling stroke, enhancing coherences in the position representation. Second, the square frequency of the trap exhibits as well a nonmonotonic behavior characterized, acquiring transient negative values associated with a purely imaginary frequency, e.g., the inversion of the trap into an anti-trap. Such inversions are also common to counterdiabatic driving for unitary processes whenever the duration of the process is comparable to $\om_0^{-1}$ \cite{Chen10}. While the inversion of the trap is not free from technical difficulties, its realization has  been facilitated by the development of the painting potential technique and the use of digital micromirror devices \cite{amico2020} as suggested in Ref. \cite{delcampo12}.

%%%%%%%%%%%%%%%%%%%%%%%%%%%%%%%%%%%%%%%%%%%%%%%%%%%%%%%%%%%%%%%%
 \section{Summary and conclusions}

 We have introduced a universal scheme to design shortcuts to adiabaticity in open quantum systems, interacting with an environment. This scheme provides the generalization of counterdiabatic driving \cite{Demirplak03,Demirplak05}, also known as transitionless quantum driving \cite{Berry09},  to open quantum systems. It is based on first  prescribing a target trajectory for the evolution of the system, and then determining the required auxiliary Hamiltonian terms and dissipators that generate it. 
 
 The resulting dynamics admits different physical realizations. It can be associated with a driven system in the presence of balanced gain and loss, a scenario that occurs naturally, e.g., in $\mathpzc{PT}$-symmetric quantum mechanics. Alternatively, it can be implemented via a non-Markovian evolution in which the equation governing the dynamics takes a generalized Lindblad-like form. The latter is readily accessible in a variety of platforms---including trapped ions, Rydberg atoms, and superconducting qubits, among other examples---by using, e.g., digital quantum simulation techniques  \cite{Lloyd96,Barreiro11,Muller12,Georgescu14,Sweke16}. Our formalism thus enables to engineer superadiabatic open processes to speed up, i.e., heating, cooling, and isothermal strokes. 
 
 We have applied this framework to the engineering of strokes in an open two-level system. In addition, we have provided an experimentally-friendly protocol for the the controlled thermalization of a driven quantum oscillator, that can be implemented with current technology in trapped ions and ultracold gases. The framework  introduced here should find broad applications in quantum thermodynamics, and more generally, in quantum technologies requiring the fast control of an open system embedded in an environment. 

%%%%%%%%%%%%%%%%%%%%%%%%%%%%%%%%%%%%%%%%%%%%%%%%%%%%%%%%%%%%%%%%
\textit{Acknowledgements.---}We would like to thank Tapio Ala-Nissila, L\'eonce Dupays, and Jack J. Mayo for comments on the manuscript. This work is supported by ID2019-109007GA-I00. Further support by the Academy of Finland's Center of Excellence program QTF Project 312057 (to S.A.) is acknowledged. A.T.R. also acknowledges support by the QTF, Aalto University's AScI Visiting Professor Fund, and Sharif University of Technology's Office of Vice President for Research and Technology.  

%%%%%%%%%%%%%%%%%%%%%%%%%%%%%%%%%%%%%%%%%%%%%%%%%%%%%%%%%%%%%%%%
%\onecolumngrid
%\newpage
\appendix
\section{Lindblad-like master equation}

In this section, we verify that the dissipator
 \beqa
 \label{eqdiss}
\mathpzc{D}(\varrho)= \textstyle{\sum_{mn}}\gamma_{mn}\big({L}_{mn}\varrho {L}_{mn}^\dag-\frac{1}{2}\{{L}_{mn}^\dag {L}_{mn},\varrho\}\big),\nonumber\\
 \eeqa
 with the choice of the time-dependent Lindblad operators and rates given in the main text, 
  satisfies the identity
 \beqa
 \label{idtoprove}
 \mathpzc{D}(\varrho)= \textstyle{\sum_m} \partial_{t}\lambda_m(t) |m_t\ra\la m_t|.
 \eeqa
Employing the explicit form of ${L}_{mn}$ in Eq. (\ref{eqdiss}) one finds
 \beqa
  \mathpzc{D}(\varrho)
= \textstyle{\sum_{mn}} \frac{\partial_{t}\lambda_m(t)}{r}(|m_t\ra\la m_t|-|n_t\ra\la n_t|).
 \eeqa
Noting that $\sum_{n=1}^r1=r={\rm rank}(\varrho)$ and $\sum_{m=1}^r\partial_{t}\lambda_m(t)=\partial_{t}\tr[\varrho]$, it follows that
 \beqa
  \mathpzc{D}(\varrho)=& \textstyle{\sum_m} \partial_{t}\lambda_m(t)|m_t\ra\la m_t|\nonumber \\
 & -(1/r)\partial_{t}\tr[\varrho]\,\sum_n|n_t\ra\la n_t|.
\eeqa
As the second term on the RHS vanishes identically for a norm-preserving evolution, this completes the proof of Eq. (\ref{idtoprove}).

% \section{Quantum speed limit for STA in open quantum processes}

%%%%%%%%%%%%%%%%%%%%%%%%%%%%%%%%%%%%%%%%%%%%%%%%%%%%%%%%%%%%%%%%
\section{Lindblad operators for arbitrary strokes in two-level systems}
 
Consider the trajectory described by the instantaneous thermal state of a two-level system
\beqa
\varrho(t)= \sum_{\alpha=\pm} \frac{e^{\alpha\frac{\beta}{2}\sqrt{\Om^2+\Delta^2}}}{2\cosh[\frac{\beta}{2}\sqrt{\Om^2+\Delta^2}]}|\alpha_t\ra\la \alpha_t|,
\eeqa
 where $\beta$, $\Delta$, and $\Om$ are time-dependent. The  Lindblad operators are ${L}_{+-}=|+\ra\la -|$ and ${L}_{-+}=|-\ra\la +|$, as in Eq. (25) in the main text, with rates
\beqa
\gamma_{+-}(t)&=&\frac{\Delta^2 \partial_{t}\beta +\Omega\left(\Omega \partial_{t}\beta +\beta\partial_{t}\Omega \right)+\beta \Delta\partial_{t}\Delta }
{2 \sqrt{\Delta^2+\Omega^2} \left(e^{-\beta\sqrt{\Delta^2+\Omega^2}}+1\right)}, \nonumber \\
\gamma_{-+}(t)&=&-\frac{\Delta^2 \partial_{t}\beta +\Omega\left(\Omega\partial_{t}\beta +\beta \partial_{t}\Omega \right)+\beta\Delta
\partial_{t}\Delta }{2 \sqrt{\Delta^2+\Omega^2}\left(e^{\beta\sqrt{\Delta^2+\Omega^2}}+1\right)}.\nonumber
\eeqa

\section{Thermalization of a quantum oscillator \label{appC}}
We provide details to establish the equivalence of the different master equations for the fast thermalization of a quantum oscillator. To do this, we use the coordinate representation. The thermal state of a harmonic oscillator is known to be described by a Gaussian density matrix,    
\begin{equation}\label{eq:sigmaX}
\varrho (x,x',t) {=} \bra{x} \varrho(t) \ket{x'} {=} N_t e^{-A_t (x^2 + x'^2) - 2C_t x x'},
\end{equation}
with normalization constant  $N_t = \sqrt{2(A_t + C_t)/\pi}$.
The real parameters $A_t =k^2_t (1+u_t^2) / (2 (1-u_t^2)) $ and $C_t = - k_t^2 u_t / (1-u^2_t)$ follow from the inverse length $k_t = \sqrt{m\omega_t/\hbar}$ and the normalization factor $N_t = k_t \sqrt{(1- u_t)/ (\pi (1+u_t))}$.
This gives the coordinate representation of the dissipator  \eqref{ME2} as
\beqa
& & \bra{x}  \tilde{\mathpzc{D}}_{\textsc{cd}}(\tilde{\varrho})\ket{x'} =  \frac{\partial \varrho(x,x',t)}{\partial t} e^{i \frac{m \alpha_t}{2 \hbar} (x^2 - x'^2)}\nonumber \\
& &= \Big( \frac{\dot{N}_t}{N_t} {-} \dot{A}_t (x^2 {+} x'^2) {-} 2 \dot{C}_t x x'\Big) \tilde{\varrho}(x,x',t)  .
\eeqa

Given the explicit form of $N_t$, choosing $\Omega_t = - \frac{\dot{N}_t}{N_t}$, as in Eq.~\eqref{assumption},  leads to $\dot{A}_t + 2\Omega_t A_t =  -\dot{C}_t - 2\Omega_t C_t \equiv \gamma_t$. The latter corresponds to a dephasing strength, and allows recasting the dissipator as 
 \begin{eqnarray} \label{eq1}
\bra{x}   \tilde{\mathpzc{D}}_{\textsc{cd}}(\tilde{\varrho}) \ket{x'} =\Big[ && \Omega_t \Big(2 A (x^2 {+} x'^2) {+} 4 C x x' {-}1\Big) \nonumber \\
&&- \gamma_t (x-x')^2 \Big] \tilde{\varrho}(x,x',t) .
 \end{eqnarray} 
We wish to rewrite this last expression in operator form. To that end we note that $[H_0,\varrho(t)]=0$ and thus $[UH_0U^\dagger,\tilde{\varrho}(t)]=0$, whence it follows that 
\beqa
[\{\hat{x},\hat{p}\},\tilde{\varrho}]=\frac{2}{\alpha}\left[\frac{\hat{p}^2}{2m}+\frac{1}{2}m(\om_t^2+\alpha_t^2)\hat{x}^2,\tilde{\varrho}\right].\nonumber\\
\eeqa
Explicit computation using the coordinate representation of the trajectory, 
$\tilde{\varrho}(x,x',t)  = \varrho(x,x',t) e^{i\frac{m \alpha_t}{2 \hbar}(x^2 - x'^2)}$, 
yields
\beqa
&\bra{x}[\{\hat{x},\hat{p}\},\tilde{\varrho}(t)] \ket{x'} =2m\alpha_t(x^2 - x'^2)\tilde{\varrho}(x,x',t)\nonumber \\
&+2 i \hbar (2 A (x^2 + x'^2) + 4 C x x' -1)\tilde{\varrho}(x,x',t).\nonumber \\
\eeqa
As a result, the dissipator admits the operator form of the dissipator given in the main text, Eq. \eqref{Dcd_HO}. 

%\end{widetext}
%\onecolumngrid


\begin{thebibliography}{72}
\providecommand{\natexlab}[1]{#1}
\providecommand{\url}[1]{\texttt{#1}}
\expandafter\ifx\csname urlstyle\endcsname\relax
  \providecommand{\doi}[1]{doi: #1}\else
  \providecommand{\doi}{doi: \begingroup \urlstyle{rm}\Url}\fi

\bibitem[Torrontegui et~al.(2013)Torrontegui, Ib\'a{\~n}ez, Mart\'inez-Garaot,
  Modugno, del Campo, Gu\'ery-Odelin, Ruschhaupt, Chen, and
  Muga]{Torrontegui13}
Erik Torrontegui, Sara Ib\'a{\~n}ez, Sofia Mart\'inez-Garaot, Michele Modugno,
  Adolfo del Campo, David Gu\'ery-Odelin, Andreas Ruschhaupt, Xi~Chen, and
  Juan~Gonzalo Muga.
\newblock Chapter 2: Shortcuts to adiabaticity.
\newblock In Ennio Arimondo, Paul~R. Berman, and Chun~C. Lin, editors,
  \emph{Advances in Atomic, Molecular, and Optical Physics}, Vol.~62, pp.
  117. Academic Press, 2013.
\newblock \doi{10.1016/B978-0-12-408090-4.00002-5}.

\bibitem[del Campo and Kim(2019)]{delcampo19}
Adolfo del Campo and Kihwan Kim.
\newblock Focus on shortcuts to adiabaticity.
\newblock \emph{New J. Phys.}, 21:050201, 2019.
\newblock \doi{10.1088/1367-2630/ab1437}.

\bibitem[Gu\'ery-Odelin et~al.(2019)Gu\'ery-Odelin, Ruschhaupt, Kiely,
  Torrontegui, Mart\'{\i}nez-Garaot, and Muga]{guery-odelin2019}
D.~Gu\'ery-Odelin, A.~Ruschhaupt, A.~Kiely, E.~Torrontegui,
  S.~Mart\'{\i}nez-Garaot, and J.~G. Muga.
\newblock Shortcuts to adiabaticity: Concepts, methods, and applications.
\newblock \emph{Rev. Mod. Phys.}, 91:045001, 2019.
\newblock \doi{10.1103/RevModPhys.91.045001}.

\bibitem[Chen et~al.(2010)Chen, Ruschhaupt, Schmidt, del Campo, Gu\'ery-Odelin,
  and Muga]{Chen10}
Xi~Chen, A.~Ruschhaupt, S.~Schmidt, A.~del Campo, D.~Gu\'ery-Odelin, and J.~G.
  Muga.
\newblock Fast optimal frictionless atom cooling in harmonic traps: Shortcut to
  adiabaticity.
\newblock \emph{Phys. Rev. Lett.}, 104:063002, 2010.
\newblock \doi{10.1103/PhysRevLett.104.063002}.

\bibitem[Demirplak and Rice(2003)]{Demirplak03}
Mustafa Demirplak and Stuart~A Rice.
\newblock Adiabatic population transfer with control fields.
\newblock \emph{J. Phys. Chem. A}, 107:9937, 2003.
\newblock \doi{10.1021/jp030708a}.

\bibitem[Demirplak and Rice(2005)]{Demirplak05}
Mustafa Demirplak and Stuart~A Rice.
\newblock Assisted adiabatic passage revisited.
\newblock \emph{J. Phys. Chem. B}, 109:6838, 2005.
\newblock \doi{10.1021/jp040647w}.

\bibitem[Deng et~al.(2013)Deng, Wang, Liu, H\"anggi, and Gong]{Deng13}
Jiawen Deng, Qing-Hai Wang, Zhihao Liu, Peter H\"anggi, and Jiangbin Gong.
\newblock Boosting work characteristics and overall heat-engine performance via
  shortcuts to adiabaticity: Quantum and classical systems.
\newblock \emph{Phys. Rev. E}, 88:062122, 2013.
\newblock \doi{10.1103/PhysRevE.88.062122}.

\bibitem[del Campo et~al.(2014)del Campo, Goold, and Paternostro]{delcampo14}
Adolfo del Campo, J~Goold, and M~Paternostro.
\newblock More bang for your buck: Super-adiabatic quantum engines.
\newblock \emph{Sci. Rep.}, 4:6208, 2014.
\newblock \doi{10.1038/srep06208}.

\bibitem[Funo et~al.(2017)Funo, Zhang, Chatou, Kim, Ueda, and del
  Campo]{Funo17}
Ken Funo, Jing-Ning Zhang, Cyril Chatou, Kihwan Kim, Masahito Ueda, and Adolfo
  del Campo.
\newblock Universal work fluctuations during shortcuts to adiabaticity by
  counterdiabatic driving.
\newblock \emph{Phys. Rev. Lett.}, 118:100602, 2017.
\newblock \doi{10.1103/PhysRevLett.118.100602}.

\bibitem[Schaff et~al.(2010)Schaff, Song, Vignolo, and Labeyrie]{Schaff2010}
J.-F. Schaff, X.-L. Song, P.~Vignolo, and G.~Labeyrie.
\newblock Fast optimal transition between two equilibrium states.
\newblock \emph{Phys. Rev. A}, 82:033430, 2010.
\newblock \doi{10.1103/PhysRevA.82.033430}.

\bibitem[Schaff et~al.(2011)Schaff, Song, Capuzzi, Vignolo, and
  Labeyrie]{Schaff2011a}
J.-F. Schaff, X.-L. Song, P.~Capuzzi, P.~Vignolo, and G.~Labeyrie.
\newblock Shortcut to adiabaticity for an interacting Bose-Einstein condensate.
\newblock \emph{Europhys. Lett.}, 93:23001, 2011.
\newblock \doi{10.1209/0295-5075/93/23001}.

\bibitem[Bason et~al.(2012)Bason, Viteau, Malossi, Huillery, Arimondo, Fazio,
  Giovannetti, Mannella, and Morsch]{Bason12}
M.~G. Bason, M.~Viteau, N.~Malossi, P.~Huillery, E.~Arimondo, R.~Fazio,
  V.~Giovannetti, R.~Mannella, and O.~Morsch.
\newblock High-fidelity quantum driving.
\newblock \emph{Nat. Phys.}, 8:147, 2012.
\newblock \doi{10.1038/nphys2170}.

\bibitem[Rohringer et~al.(2015)Rohringer, Fischer, Steiner, Mazets,
  Schmiedmayer, and Trupke]{Rohringer15}
W.~Rohringer, D.~Fischer, F.~Steiner, I.~E. Mazets, J.~Schmiedmayer, and
  M.~Trupke.
\newblock {Non-equilibrium scale invariance and shortcuts to adiabaticity in a
  one-dimensional Bose gas}.
\newblock \emph{Sci. Rep.}, 5:9820, 2015.
\newblock \doi{10.1038/srep09820}.

\bibitem[Deng et~al.(2018{\natexlab{a}})Deng, Diao, Yu, del Campo, and
  Wu]{Deng18pra}
Shujin Deng, Pengpeng Diao, Qianli Yu, Adolfo del Campo, and Haibin Wu.
\newblock Shortcuts to adiabaticity in the strongly coupled regime:
  Nonadiabatic control of a unitary fermi gas.
\newblock \emph{Phys. Rev. A}, 97:013628, 2018{\natexlab{a}}.
\newblock \doi{10.1103/PhysRevA.97.013628}.

\bibitem[Deng et~al.(2018{\natexlab{b}})Deng, Chenu, Diao, Li, Yu, Coulamy, del
  Campo, and Wu]{Deng18Sci}
Shujin Deng, Aur{\'e}lia Chenu, Pengpeng Diao, Fang Li, Shi Yu, Ivan Coulamy,
  Adolfo del Campo, and Haibin Wu.
\newblock Superadiabatic quantum friction suppression in finite-time
  thermodynamics.
\newblock \emph{Sci. Adv.}, 4:eaar5909,
  2018{\natexlab{b}}.
\newblock \doi{10.1126/sciadv.aar5909}.

\bibitem[Diao et~al.(2018)Diao, Deng, Li, Yu, Chenu, del Campo, and Wu]{Diao18}
Pengpeng Diao, Shujin Deng, Fang Li, Shi Yu, Aur{\'{e}}lia Chenu, Adolfo del
  Campo, and Haibin Wu.
\newblock Shortcuts to adiabaticity in fermi gases.
\newblock \emph{New J. Phys.}, 20:105004, oct
  2018.
\newblock \doi{10.1088/1367-2630/aae45e}.

\bibitem[Zhang et~al.(2013)Zhang, Shim, Niemeyer, Taniguchi, Teraji, Abe,
  Onoda, Yamamoto, Ohshima, Isoya, and Suter]{Zhang13}
Jingfu Zhang, Jeong~Hyun Shim, Ingo Niemeyer, T.~Taniguchi, T.~Teraji, H.~Abe,
  S.~Onoda, T.~Yamamoto, T.~Ohshima, J.~Isoya, and Dieter Suter.
\newblock Experimental implementation of assisted quantum adiabatic passage in
  a single spin.
\newblock \emph{Phys. Rev. Lett.}, 110:240501, 2013.
\newblock \doi{10.1103/PhysRevLett.110.240501}.

\bibitem[K\"olbl et~al.(2019)K\"olbl, Barfuss, Kasperczyk, Thiel, Clerk,
  Ribeiro, and Maletinsky]{Koelbl19}
J.~K\"olbl, A.~Barfuss, M.~S. Kasperczyk, L.~Thiel, A.~A. Clerk, H.~Ribeiro,
  and P.~Maletinsky.
\newblock Initialization of single spin dressed states using shortcuts to
  adiabaticity.
\newblock \emph{Phys. Rev. Lett.}, 122:090502, 2019.
\newblock \doi{10.1103/PhysRevLett.122.090502}.

\bibitem[An et~al.(2016)An, Lv, del Campo, and Kim]{An16}
Shuoming An, Dingshun Lv, Adolfo del Campo, and Kihwan Kim.
\newblock Shortcuts to adiabaticity by counterdiabatic driving for trapped-ion
  displacement in phase space.
\newblock \emph{Nat. Commun.}, 7:12999, 2016.
\newblock \doi{10.1038/ncomms12999}.

\bibitem[Wang et~al.(2018)Wang, Zhang, Xiang, Jia, Duan, Cai, Gong, Zong, Wu,
  Wu, Sun, Yin, and Guo]{Wang18}
Tenghui Wang, Zhenxing Zhang, Liang Xiang, Zhilong Jia, Peng Duan, Weizhou Cai,
  Zhihao Gong, Zhiwen Zong, Mengmeng Wu, Jianlan Wu, Luyan Sun, Yi~Yin, and
  Guoping Guo.
\newblock The experimental realization of high-fidelity
  `shortcut-to-adiabaticity' quantum gates in a superconducting xmon qubit.
\newblock \emph{New J. Phys.}, 20:065003, 2018.
\newblock \doi{10.1088/1367-2630/aac9e7}.

\bibitem[Wang et~al.(2019)Wang, Zhang, Xiang, Jia, Duan, Zong, Sun, Dong, Wu,
  Yin, and Guo]{Wang19}
Tenghui Wang, Zhenxing Zhang, Liang Xiang, Zhilong Jia, Peng Duan, Zhiwen Zong,
  Zhenhai Sun, Zhangjingzi Dong, Jianlan Wu, Yi~Yin, and Guoping Guo.
\newblock Experimental realization of a fast controlled-$Z$ gate via a shortcut
  to adiabaticity.
\newblock \emph{Phys. Rev. Applied}, 11:034030, 2019.
\newblock \doi{10.1103/PhysRevApplied.11.034030}.

\bibitem[{Breuer} and {Petruccione}(2007)]{BreuerBook}
H.-P. {Breuer} and P.~{Petruccione}.
\newblock \emph{The Theory of Open Quantum Systems}.
\newblock Oxford University Press, Oxford, 2007.

\bibitem[Plenio and Knight(1998)]{PlenioKnight98}
M.~B. Plenio and P.~L. Knight.
\newblock The quantum-jump approach to dissipative dynamics in quantum optics.
\newblock \emph{Rev. Mod. Phys.}, 70:101, 1998.
\newblock \doi{10.1103/RevModPhys.70.101}.

\bibitem[Bender and Boettcher(1998)]{Bender98}
Carl~M. Bender and Stefan Boettcher.
\newblock Real spectra in non-Hermitian Hamiltonians having
  $\mathpzc{P}\mathpzc{T}$ symmetry.
\newblock \emph{Phys. Rev. Lett.}, 80:5243, 1998.
\newblock \doi{10.1103/PhysRevLett.80.5243}.

\bibitem[R{\"u}ter et~al.(2010)R{\"u}ter, Makris, El-Ganainy, Christodoulides,
  Segev, and Kip]{Ruter2010}
Christian~E. R{\"u}ter, Konstantinos~G. Makris, Ramy El-Ganainy, Demetrios~N.
  Christodoulides, Mordechai Segev, and Detlef Kip.
\newblock Observation of parity-time symmetry in optics.
\newblock \emph{Nat. Phys.}, 6:192, 2010.
\newblock \doi{10.1038/nphys1515}.

\bibitem[Regensburger et~al.(2012)Regensburger, Bersch, Miri, Onishchukov,
  Christodoulides, and Peschel]{Regensburger2012}
Alois Regensburger, Christoph Bersch, Mohammad-Ali Miri, Georgy Onishchukov,
  Demetrios~N. Christodoulides, and Ulf Peschel.
\newblock Parity-time synthetic photonic lattices.
\newblock \emph{Nature (London)}, 488:167, 2012.
\newblock \doi{10.1038/nature11298}.

\bibitem[Feng et~al.(2012)Feng, Xu, Fegadolli, Lu, Oliveira, Almeida, Chen, and
  Scherer]{Feng2012}
Liang Feng, Ye-Long Xu, William~S. Fegadolli, Ming-Hui Lu, Jos{\'e} E.~B.
  Oliveira, Vilson~R. Almeida, Yan-Feng Chen, and Axel Scherer.
\newblock Experimental demonstration of a unidirectional reflectionless
  parity-time metamaterial at optical frequencies.
\newblock \emph{Nat. Mater.}, 12:108, 2012.
\newblock \doi{10.1038/nmat3495}.

\bibitem[Peng et~al.(2014)Peng, {\"O}zdemir, Lei, Monifi, Gianfreda, Long, Fan,
  Nori, Bender, and Yang]{Peng2014}
Bo~Peng, Sahin~Kaya {\"O}zdemir, Fuchuan Lei, Faraz Monifi, Mariagiovanna
  Gianfreda, Gui~Lu Long, Shanhui Fan, Franco Nori, Carl~M. Bender, and Lan
  Yang.
\newblock Parity-time-symmetric whispering-gallery microcavities.
\newblock \emph{Nat. Phys.}, 10:394, 2014.
\newblock \doi{10.1038/nphys2927}.

\bibitem[Zhen et~al.(2015)Zhen, Hsu, Igarashi, Lu, Kaminer, Pick, Chua,
  Joannopoulos, and {Solja\v{c}i\'{c}}]{Zhen2015}
Bo~Zhen, Chia~Wei Hsu, Yuichi Igarashi, Ling Lu, Ido Kaminer, Adi Pick,
  Song-Liang Chua, John~D. Joannopoulos, and Marin {Solja\v{c}i\'{c}}.
\newblock Spawning rings of exceptional points out of dirac cones.
\newblock \emph{Nature (London)}, 525:354, 2015.
\newblock \doi{10.1038/nature14889}.

\bibitem[Ruschhaupt et~al.(2012)Ruschhaupt, Chen, Alonso, and
  Muga]{ruschhaupt2012}
A~Ruschhaupt, Xi~Chen, D~Alonso, and J~G Muga.
\newblock Optimally robust shortcuts to population inversion in two-level
  quantum systems.
\newblock \emph{New J. Phys.}, 14: 093040, 2012.
\newblock \doi{10.1088/1367-2630/14/9/093040}.

\bibitem[Kiely and Ruschhaupt(2014)]{kiely2014}
Anthony Kiely and Andreas Ruschhaupt.
\newblock Inhibiting unwanted transitions in population transfer in two-and
  three-level quantum systems.
\newblock \emph{J. Phys. B: At. Mol. Opt. Phys.}, 47:115501, 2014.
\newblock \doi{10.1088/0953-4075/47/11/115501}.

\bibitem[Lidar et~al.(1998)Lidar, Chuang, and Whaley]{Lidar98}
D.~A. Lidar, I.~L. Chuang, and K.~B. Whaley.
\newblock Decoherence-free subspaces for quantum computation.
\newblock \emph{Phys. Rev. Lett.}, 81:2594, 1998.
\newblock \doi{10.1103/PhysRevLett.81.2594}.

\bibitem[Wu et~al.(2017)Wu, Huang, Li, and Yi]{wu2017}
S.~L. Wu, X.~L. Huang, H.~Li, and X.~X. Yi.
\newblock Adiabatic evolution of decoherence-free subspaces and its shortcuts.
\newblock \emph{Phys. Rev. A}, 96:042104, 2017.
\newblock \doi{10.1103/PhysRevA.96.042104}.

\bibitem[Levy et~al.(2018)Levy, Kiely, Muga, Kosloff, and
  Torrontegui]{levy2018}
Amikam Levy, A.~Kiely, J.~G. Muga, R.~Kosloff, and E.~Torrontegui.
\newblock Noise resistant quantum control using dynamical invariants.
\newblock \emph{New J. Phys.}, 20:025006, 2018.
\newblock \doi{10.1088/1367-2630/aaa9e5}.

\bibitem[{Boyd} et~al.(2018){Boyd}, {Patra}, {Jarzynski}, and
  {Crutchfield}]{Boyd18}
A.~B. {Boyd}, A.~{Patra}, C.~{Jarzynski}, and J.~P. {Crutchfield}.
\newblock Shortcuts to thermodynamic computing: The cost of fast and faithful
  erasure.
\newblock \emph{arXiv:1812.11241}, 2018.
\newblock URL \url{https://arxiv.org/abs/1812.11241}.

\bibitem[{del Campo} et~al.({2018}){del Campo}, {Chenu}, {Deng}, and
  {Wu}]{delcampo18}
Adolfo {del Campo}, Aur{\'e}lia {Chenu}, Shujin {Deng}, and Haibin {Wu}.
\newblock \emph{Friction-Free Quantum Machines}.
\newblock In F. Binder, L. Correa, C. Gogolin, J. Anders, and G. Adesso, editors,
  \emph{Thermodynamics in the Quantum Regime}, pp. 127.
\newblock {Springer International Publishing}, {Cham}, {2018}.
\newblock \doi{10.1007/978-3-319-99046-0_5}.


\bibitem[Ib\'a\~nez et~al.(2011)Ib\'a\~nez, Mart\'{\i}nez-Garaot, Chen,
  Torrontegui, and Muga]{ibanez2011}
S.~Ib\'a\~nez, S.~Mart\'{\i}nez-Garaot, Xi~Chen, E.~Torrontegui, and J.~G.
  Muga.
\newblock Shortcuts to adiabaticity for non-Hermitian systems.
\newblock \emph{Phys. Rev. A}, 84:023415, 2011.
\newblock \doi{10.1103/PhysRevA.84.023415}.

\bibitem[Li et~al.(2017)Li, Chen, Peng, and Qi]{li2017}
Guan-Qiang Li, Guang-De Chen, Ping Peng, and Wei Qi.
\newblock Non-Hermitian shortcut to adiabaticity of two-and three-level systems
  with gain and loss.
\newblock \emph{Eur. Phys. J. D}, 71:1, 2017.
\newblock \doi{10.1140/epjd/e2016-70525-6}.

\bibitem[Chen et~al.(2018)Chen, Wu, Huang, Song, Xia, and Zheng]{chen2018}
Ye-Hong Chen, Qi-Cheng Wu, Bi-Hua Huang, Jie Song, Yan Xia, and Shi-Biao Zheng.
\newblock Improving shortcuts to non-Hermitian adiabaticity for fast population
  transfer in open quantum systems.
\newblock \emph{Ann. Phys. (Berlin)}, 530: 1700247, 2018.
\newblock \doi{10.1002/andp.201700247}.

\bibitem[Impens and Gu{\'e}ry-Odelin(2019)]{impens2019}
Fran{\c{c}}ois Impens and David Gu{\'e}ry-Odelin.
\newblock Fast quantum control in dissipative systems using dissipationless
  solutions.
\newblock \emph{Sci. Rep.}, 9:1, 2019.
\newblock \doi{10.1038/s41598-019-39731-z}.

\bibitem[Vacanti et~al.(2014)Vacanti, Fazio, Montangero, Palma, Paternostro,
  and Vedral]{Vacanti14}
G.~Vacanti, R.~Fazio, S.~Montangero, G.~M. Palma, M.~Paternostro, and V.~Vedral.
\newblock Transitionless quantum driving in open quantum systems.
\newblock \emph{New J. Phys.}, 16:053017, 2014.
\newblock \doi{10.1088/1367-2630/16/5/053017}.

\bibitem[Sarandy and Lidar(2005)]{sarandy2005}
M.~S. Sarandy and D.~A. Lidar.
\newblock Adiabatic approximation in open quantum systems.
\newblock \emph{Phys. Rev. A}, 71:012331, 2005.
\newblock \doi{10.1103/PhysRevA.71.012331}.

\bibitem[Dann et~al.(2019)Dann, Tobalina, and Kosloff]{Dann19}
Roie Dann, Ander Tobalina, and Ronnie Kosloff.
\newblock Shortcut to equilibration of an open quantum system.
\newblock \emph{Phys. Rev. Lett.}, 122:250402, 2019.
\newblock \doi{10.1103/PhysRevLett.122.250402}.

\bibitem[Dupays et~al.(2020)Dupays, Egusquiza, del Campo, and Chenu]{dupays20}
L.~Dupays, I.~L. Egusquiza, A.~del Campo, and A.~Chenu.
\newblock Superadiabatic thermalization of a quantum oscillator by engineered
  dephasing.
\newblock \emph{Phys. Rev. Research}, 2:033178, 2020.
\newblock \doi{10.1103/PhysRevResearch.2.033178}.

\bibitem[{Villazon} et~al.(2019){Villazon}, {Polkovnikov}, and
  {Chandran}]{Villazon19}
Tamiro {Villazon}, Anatoli {Polkovnikov}, and Anushya {Chandran}.
\newblock {Swift heat transfer by fast-forward driving in open quantum
  systems}.
\newblock \emph{Phys. Rev. A}, 100:012126, 2019.
\newblock \doi{10.1103/PhysRevA.100.012126}.

\bibitem[Pancotti et~al.(2019)Pancotti, Scandi, Mitchison, and
  Perarnau-Llobet]{pancotti2019}
N.~Pancotti, M.~Scandi, M.~T. Mitchison, and M.~Perarnau-Llobet.
\newblock Speed-ups to isothermality: Enhanced quantum heat engines through
  control of the system-bath coupling.
\newblock \emph{arXiv:1911.12437}, 2019.
\newblock URL \url{https://arxiv.org/abs/1911.12437}.

\bibitem[Dupays and Chenu(2020)]{dupays2020dynamical}
L.~Dupays and A.~Chenu.
\newblock Dynamical engineering of squeezed thermal states, 2020.
\newblock URL \url{https://arxiv.org/abs/2008.03307}.

\bibitem[Berry(2009)]{Berry09}
M.~V. Berry.
\newblock Transitionless quantum driving.
\newblock \emph{J. Phys. A: Math. Theor.}, 42:365303, 2009.
\newblock \doi{10.1088/1751-8113/42/36/365303}.

\bibitem[Kato(1950)]{Kato50}
Tosio Kato.
\newblock On the adiabatic theorem of quantum mechanics.
\newblock \emph{J. Phys. Soc. Jpn.}, 5:435, 1950.
\newblock \doi{10.1143/JPSJ.5.435}.

\bibitem[Avron et~al.(1987)Avron, Seiler, and Yaffe]{Avron87}
J.~E. Avron, R.~Seiler, and L.~G. Yaffe.
\newblock Adiabatic theorems and applications to the quantum Hall effect.
\newblock \emph{Commun. Math. Phys.}, 110:33, 1987.
\newblock \doi{10.1007/BF01209015}.

\bibitem[Brody and Graefe(2012)]{Brody12}
Dorje~C. Brody and Eva-Maria Graefe.
\newblock Mixed-state evolution in the presence of gain and loss.
\newblock \emph{Phys. Rev. Lett.}, 109:230405, 2012.
\newblock \doi{10.1103/PhysRevLett.109.230405}.

\bibitem[Gong and Wang(2013)]{Gong13}
Jiangbin Gong and Qing-Hai Wang.
\newblock Time-dependent $\mathcal{PT}$-symmetric quantum mechanics.
\newblock \emph{J. Phys. A: Math. Theor.}, 46:485302, 2013.
\newblock \doi{10.1088/1751-8113/46/48/485302}.

\bibitem[{Alipour} et~al.(2020){Alipour}, {Rezakhani}, {Babu}, {M{\o}lmer},
  {M\"{o}tt\"{o}nen}, and {Ala-Nissila}]{ULL}
S.~{Alipour}, A.~T. {Rezakhani}, A.~P. {Babu}, K.~{M{\o}lmer},
  M.~{M\"{o}tt\"{o}nen}, and T.~{Ala-Nissila}.
\newblock {Correlation-Picture Approach to Open-Quantum-System Dynamics}.
\newblock \emph{arXiv:1903.03861 (to appear in PRX)}, 2020.
\newblock URL \url{https://arxiv.org/abs/1903.03861}.

\bibitem[{Funo} et~al.(2019){Funo}, {Shiraishi}, and
  {Saito}]{Funo:QSL-OpenSystem}
K.~{Funo}, N.~{Shiraishi}, and K.~{Saito}.
\newblock {Speed limit for open quantum systems}.
\newblock \emph{New J. Phys.}, 21:013006, 2019.
\newblock \doi{10.1088/1367-2630/aaf9f5}.

\bibitem[{Rezakhani} et~al.(2010){Rezakhani}, {Abasto}, {Lidar}, and
  {Zanardi}]{Rezakhani-Abasto}
A.~T. {Rezakhani}, D.~F. {Abasto}, D.~A. {Lidar}, and P.~{Zanardi}.
\newblock {Intrinsic geometry of quantum adiabatic evolution and quantum phase
  transitions}.
\newblock \emph{Phys. Rev. A}, 82:012321, 2010.
\newblock \doi{10.1103/PhysRevA.82.012321}.

\bibitem[Alipour and Rezakhani(2015)]{Alipour-convexity}
S.~Alipour and A.~T. Rezakhani.
\newblock Extended convexity of quantum Fisher information in quantum
  metrology.
\newblock \emph{Phys. Rev. A}, 91:042104, 2015.
\newblock \doi{10.1103/PhysRevA.91.042104}.

\bibitem[Alipour et~al.(2016)Alipour, Benatti, Bakhshinezhad, Afsary,
  Marcantoni, and Rezakhani]{SciRep:Alipour-corr}
S.~Alipour, F.~Benatti, F.~Bakhshinezhad, M.~Afsary, S.~Marcantoni, and A.~T.
  Rezakhani.
\newblock Correlations in quantum thermodynamics: Heat, work, and entropy
  production.
\newblock \emph{Sci. Rep.}, 6:35568, 2016.
\newblock \doi{10.1103/PhysRevX.4.031042}.

\bibitem[Rendell and Rajagopal(2003)]{Rajagopal}
R.~W. Rendell and A.~K. Rajagopal.
\newblock Revivals and entanglement from initially entangled mixed states of a
  damped jaynes-cummings model.
\newblock \emph{Phys. Rev. A}, 67:062110, 2003.
\newblock \doi{10.1103/PhysRevA.67.062110}.

\bibitem[Carmichael(1993)]{Book:Carmichael}
H.~Carmichael.
\newblock \emph{An Open Systems Approach to Quantum Optics}.
\newblock Springer, Berlin, 1993.

\bibitem[Muga et~al.(2010)Muga, Chen, Ib{\'{a}}{\~{n}}ez, Lizuain, and
  Ruschhaupt]{Muga2010}
J.~G. Muga, X.~Chen, S.~Ib{\'{a}}{\~{n}}ez, I.~Lizuain, and A.~Ruschhaupt.
\newblock Transitionless quantum drivings for the harmonic oscillator.
\newblock \emph{J. Phys. B: At. Mol. Opt. Phys.},
  43: 085509, 2010.
\newblock \doi{10.1088/0953-4075/43/8/085509}.

\bibitem[Jarzynski(2013)]{Jarzynski13}
Christopher Jarzynski.
\newblock Generating shortcuts to adiabaticity in quantum and classical
  dynamics.
\newblock \emph{Phys. Rev. A}, 88:040101, 2013.
\newblock \doi{10.1103/PhysRevA.88.040101}.

\bibitem[del Campo(2013)]{delcampo-13}
Adolfo del Campo.
\newblock Shortcuts to adiabaticity by counterdiabatic driving.
\newblock \emph{Phys. Rev. Lett.}, 111:100502, 2013.
\newblock \doi{10.1103/PhysRevLett.111.100502}.

\bibitem[Ib\'a\~nez et~al.(2012)Ib\'a\~nez, Chen, Torrontegui, Muga, and
  Ruschhaupt]{Ibanez12}
S.~Ib\'a\~nez, Xi~Chen, E.~Torrontegui, J.~G. Muga, and A.~Ruschhaupt.
\newblock Multiple Schr\"odinger pictures and dynamics in shortcuts to
  adiabaticity.
\newblock \emph{Phys. Rev. Lett.}, 109:100403, 2012.
\newblock \doi{10.1103/PhysRevLett.109.100403}.

\bibitem[Smith et~al.(2018)Smith, Lu, An, Zhang, Zhang, Gong, Quan, Jarzynski,
  and Kim]{Smith18}
Andrew Smith, Yao Lu, Shuoming An, Xiang Zhang, Jing-Ning Zhang, Zongping Gong,
  H.~T. Quan, Christopher Jarzynski, and Kihwan Kim.
\newblock Verification of the quantum nonequilibrium work relation in the
  presence of decoherence.
\newblock \emph{New J. Phys.}, 20:013008, 2018.
\newblock \doi{10.1088/1367-2630/aa9cd6}.

\bibitem[et~al.(2020)]{amico2020}
L.~Amico \textit{et~al}.
\newblock Roadmap on atomtronics.
\newblock \emph{arXiv:2008.04439}, 2020.
\newblock URL \url{http://arxiv.org/abs/2008.04439}.





\bibitem[del Campo and Boshier(2012)]{delcampo12}
A.~del Campo and M.~G. Boshier.
\newblock Shortcuts to adiabaticity in a time-dependent box.
\newblock \emph{Sci. Rep.}, 2:648, 2012.
\newblock \doi{10.1038/srep00648}.

\bibitem[Lloyd(1996)]{Lloyd96}
Seth Lloyd.
\newblock Universal quantum simulators.
\newblock \emph{Science}, 273:1073, 1996.
\newblock \doi{10.1126/science.273.5278.1073}.

\bibitem[Barreiro et~al.(2011)Barreiro, M{\"u}ller, Schindler, Nigg, Monz,
  Chwalla, Hennrich, Roos, Zoller, and Blatt]{Barreiro11}
Julio~T. Barreiro, Markus M{\"u}ller, Philipp Schindler, Daniel Nigg, Thomas
  Monz, Michael Chwalla, Markus Hennrich, Christian~F. Roos, Peter Zoller, and
  Rainer Blatt.
\newblock An open-system quantum simulator with trapped ions.
\newblock \emph{Nature (London)}, 470:486, 2011.
\newblock \doi{10.1038/nature09801}.

\bibitem[Müller et~al.(2012)Müller, Diehl, Pupillo, and Zoller]{Muller12}
Markus M\"{u}ller, Sebastian Diehl, Guido Pupillo, and Peter Zoller.
\newblock Engineered open systems and quantum simulations with atoms and ions.
\newblock In Paul Berman, Ennio Arimondo, and Chun Lin, editors, \emph{Advances
  in Atomic, Molecular, and Optical Physics}, Vol.~61 of \emph{Advances in
  Atomic, Molecular, and Optical Physics}, pp. 1. Academic Press, 2012.
\newblock \doi{10.1016/B978-0-12-396482-3.00001-6}.

\bibitem[Georgescu et~al.(2014)Georgescu, Ashhab, and Nori]{Georgescu14}
I.~M. Georgescu, S.~Ashhab, and Franco Nori.
\newblock Quantum simulation.
\newblock \emph{Rev. Mod. Phys.}, 86:153, 2014.
\newblock \doi{10.1103/RevModPhys.86.153}.

\bibitem[Sweke et~al.(2016)Sweke, Sanz, Sinayskiy, Petruccione, and
  Solano]{Sweke16}
R.~Sweke, M.~Sanz, I.~Sinayskiy, F.~Petruccione, and E.~Solano.
\newblock Digital quantum simulation of many-body non-Markovian dynamics.
\newblock \emph{Phys. Rev. A}, 94:022317, 2016.
\newblock \doi{10.1103/PhysRevA.94.022317}.

\end{thebibliography}
\end{document}